\documentclass[twocolumn,nofootinbib,preprintnumbers,prd,superscriptaddress]{revtex4}

\usepackage{amsmath}
\usepackage{amsfonts}
\usepackage{amssymb}
\usepackage{graphicx}
\usepackage[titletoc]{appendix}
\usepackage{color}

\usepackage[rightcaption]{sidecap}
\usepackage{subfigure}
\usepackage{comment}

\usepackage{dcolumn}

\usepackage{tikz}
\usetikzlibrary{decorations.text}

\usepackage{array}
\usepackage{ctable}
\usepackage{multirow}
\usepackage{siunitx}
\usepackage{longtable}
\usepackage{tabularx}
\usepackage{booktabs}

\usepackage{hyperref}
\usepackage{cleveref}

\graphicspath{{Graphics/}}

\def\be{\begin{equation}}
\def\ee{\end{equation}}
\def\bea{\begin{eqnarray}}
\def\eea{\end{eqnarray}}

\definecolor{vividviolet}{rgb}{0.62, 0.0, 1.0}
\definecolor{amaranth}{rgb}{0.9, 0.17, 0.31}
\definecolor{palatinateblue}{rgb}{0.15, 0.23, 0.89}
\definecolor{brightpink}{rgb}{1.0, 0.0, 0.5}
\definecolor{cornflowerblue}{rgb}{0.39, 0.58, 0.93}
\definecolor{deepcarminepink}{rgb}{0.94, 0.19, 0.22}
\definecolor{radicalred}{rgb}{1.0, 0.21, 0.37}
\hypersetup{ linktoc=all,
    colorlinks, linkcolor={palatinateblue},
    citecolor={brightpink}, urlcolor={amaranth}
}

\begin{document}

\title{Entanglement entropy for spherically symmetric regular black holes}

\author{Orlando Luongo}
\email{orlando.luongo@unicam.it}
\affiliation{Universit\`a di Camerino, Via Madonna delle Carceri 9, 62032 Camerino, Italy.}
\affiliation{SUNY Polytechnic Institute, 13502 Utica, New York, USA.}
\affiliation{Istituto Nazionale di Fisica Nucleare (INFN), Sezione di Perugia, Perugia, 06123, Italy.}
\affiliation{INAF - Osservatorio Astronomico di Brera, Milano, Italy.}
\affiliation{NNLOT, Al-Farabi Kazakh National University, Al-Farabi av. 71, 050040 Almaty, Kazakhstan.}

\author{Stefano Mancini}
\email{stefano.mancini@unicam.it}
\affiliation{Universit\`a di Camerino, Via Madonna delle Carceri 9, 62032 Camerino, Italy.}
\affiliation{Istituto Nazionale di Fisica Nucleare (INFN), Sezione di Perugia, Perugia, 06123, Italy.}

\author{Paolo Pierosara}
\email{paolo.pierosara@studenti.unicam.it}
\affiliation{Universit\`a di Camerino, Via Madonna delle Carceri 9, 62032 Camerino, Italy.}

\date{\today}

\begin{abstract}
The Bardeen and Hayward spacetimes are here considered as standard configurations of spherically symmetric regular black holes. Assuming the thermodynamics of such objects to be analogous to standard black holes, we compute the island formula in the regime of small topological charge and large vacuum energy, respectively for Bardeen and Hayward spacetimes. Late and early-time domains are separately discussed, with particular emphasis on the island formations. We single out conditions under which it is not possible to find out islands at early-times and how our findings depart from the standard Schwarzschild case. Motivated by the fact that those configurations extend  Reissner-Nordstr\"{o}m and Schwarzschild-de Sitter metrics through the inclusion of regularity behavior at $r=0$, we show how the effect of regularity induces modifications on the overall entanglement entropy. Finally, the Page time is also computed, showing its asymptotic behavior. Specifically, the Page time exhibits slight departures with respect to the Schwarzschild case, more evident in the Hayward metric only. Conversely, our findings on Page time show that the Bardeen regular black hole is quite indistinguishable from the Schwarzschild case.
\end{abstract}


\maketitle
\tableofcontents

\section{Introduction}\label{introduzione}

Black holes (BHs) are some of the most significant objects in the universe. Gaining a better understanding of their physical properties could potentially reveal departures from Einstein's theory of gravity, as BHs possess  strong gravitational fields where quantum gravity effects may come into play. We are currently witnessing the dawn of a new era in precision astronomy based on black holes (BHs), as evidenced by the detection of gravitational waves \cite{LIGOScientific:2016aoc} and by the remarkable discovery of BH shadows \cite{EventHorizonTelescope:2019dse} \cite{2021NatRP...3..732V}.

Specifically, it has been established that the presence of matter that satisfies reasonable energy conditions inevitably leads to singularities, for theories that satisfy the equivalence principle, as demonstrated by the Penrose and Hawking singularity theorems \cite{Penrose:65collapse, Hawking:70sing}.

In addition, among all possible theoretical studies on BHs, the \emph{ information paradox}  represents a major issue of quantum gravity \cite{Hawking:76breakdown} and, more broadly, of general relativity and effective field theories. In particular, Hawking's radiation turns out to be  thermal, namely the entanglement entropy outside a BH  consistently increases \cite{Hawking:75particle}. This phenomenon is in contrast to quantum mechanics, where the entanglement entropy might eventually reach zero as the BH approaches the end of its evaporation. This is due to the pure states that occur at the end of evaporation.

To prompt this issue, one can investigate the Page curve \cite{Page:93information,Page:2012time}, that displays the entanglement entropy time evolution, leaving \emph{de facto} open the caveat of how the Page curve can be reproduced for the entanglement entropy of Hawking radiation. This ensures how to solve the problem of information loss quantum field theories in gravitational contexts, i.e., in curved spacetimes. Following Page's treatment, a restoration of unitarity, involving entropy decreasing after the Page time \cite{Hawking:76breakdown,Page:93information,Page:2012time}, can be found. This appears essential in conclusively solving the information paradox. Thus, a physical mechanism \emph{fueling} the Page curve to exist could represent a key to guarantee the Page process to occur. To this end,  it has been recently proposed that the Page curve arises from the effect of peculiar \emph{islands} \cite{penington:2020entanglement,Almheiri:2019bulk,Almheiri:2020semiclassical, almheiri:2019islands,Omidi:2022hyperscaling,yu2021pageDilaton,Yu:2022islandSuperradiance,Yu:2023islandGenDilaton,Geng:2020massiveIsland,Geng:2021entPhaseStruc,Geng:2020infTransf,Geng:2021inconsistency,Geng:2021infPardeSitter,Ahn:2021chg,Karananas:2021dilaton}.

{To clarify this point, we notice that the unitarity evolution of the black hole evaporation corresponds to that the entanglement entropy of Hawking radiation following the Page curve. Therefore, reproducing the Page curve for the time evolution of the black hole evaporation is an important step towards resolving the information loss paradox of black holes.

Recently, a significant progress in deducing the Page curve for the entanglement entropy of Hawking radiation, by taking into account the configuration with the islands, has been shown \cite{Hashimoto_2020,Ryu:2006holographic,Lewkowycz:2013generalized,Engelhardt:2015extremal,Hubeny:2007covariant,Faulkner:2013corrections}.

The islands are some regions, $I$, which are completely disconnected from the region, $R$, of Hawking radiation, which is assumed to be far away from black holes such that the backreaction of Hawking quanta on the spacetime geometry is negligible. The boundaries of the islands extremize the generalized entropy functional and hence they are called the extremal surfaces.

A density matrix relating to the systems of Hawking radiation is usually computed by taking the partial trace over the systems in the complementary part of the radiation region. While no assumption on the purity of the global state is made here, by using the quantum extremal surface technique, it was however found that the islands appear in the complementary region of $R$. Hence, the systems in the islands should be eliminated from those that are traced out.

According to the quantum extremal surface prescription, the entanglement entropy of Hawking radiation is obtained as the minimum value of the generalized entropy functional.

}

The subsequent strategy underlying the above prescription implies that one first extremizes the \emph{generalized entropy} to locate the extremal points. Those, indeed, indicate the island locations and therefore the entropy minimum value becomes the fine-grained entropy of radiation \cite{penington:2020entanglement,Almheiri:2019bulk,Almheiri:2020semiclassical, almheiri:2019islands}. The interest toward the concept of island increases as the same generalized entropy can be found adopting  the replica method applied to the gravitational path integral \cite{penington:2020replica,Almheiri:2020wormholes} and, moreover, the island formula can be understood by combining the AdS/BCFT correspondence and the brane world holography \cite{Callan:1994geometric,Holzhey:1994renormalized,Calabrese:2009entanglement,Chen:2020evaporation,Chen:2020pulling,akers:2020quantum,Liu:2021dynamical,Marolf:2020transcenting,Balasubramanian:2021secret,Bhattacharya:2020purification,verlinde:2020er,Chen:2020majorana,Gautason:2020page,Anegawa:2020notes,Almheiri:2020higher,schwarzschild:1999gravitational}.

As previously mentioned, BH singularities are believed to occur at a classical level but may be resolved by introducing a complete theory of quantum gravity. To address the issue of singularities at a classical level, Bardeen introduced the concept of a regular black hole (RBH) \cite{bardeen1968proceedings}. A RBH exhibits asymptotic flatness and a non-singular center in a static spherical symmetry, and it has been shown to be a genuine solution of Einstein's gravity \cite{Ayon-Beato:2000mjt}. The idea behind the RBH configuration involves a magnetic monopole source within the context of nonlinear electrodynamics \cite{ayonbeato:1998regular}. Subsequently, other models of RBHs were proposed, and notably, the Hayward solution was introduced. This solution is static, spherically symmetric, and addresses the information-loss paradox \cite{Hayward:2006nonsingular}. The use of RBHs is not only speculative; they have been assumed to model neutron stars that feature quasi-periodic oscillations. Other approaches have suggested that one cannot exclude topological charge effects and non-zero vacuum energy at $r=0$, as seen in the Hayward solution.

Since RBHs exhibit horizons, they offer a suitable setting to investigate how the island formula works. This paper focuses on describing the early and late-time approximations required to evaluate the island formula. Instead of using holographic correspondence, we demonstrate that the entanglement entropy due to Hawking radiation follows the Page curve when islands are involved in RBHs. We evaluate this using the two main spacetime configurations quoted above, namely Bardeen and Hayward metrics. Specifically, we investigate the effects of topological charge and vacuum energy in computing the island formula. We consider the overall domain where islands can arise without limiting the analysis to large or small radii, evaluating the regions of the islands for small topological and large vacuum energy contributions. After considering the early and late-time approximations, we check how the Page curves are reproduced. The results differ from the simplest Schwarzschild case due to the presence of corrective terms within the RBH metrics. Slight departures arise for the Hayward spacetime only concerning Page time, while the Bardeen solution appears quite indistinguishable from the Schwarzschild solution. Finally, we discuss the physical consequences of our approach with respect to current literature.

The paper is structured as follows. In Sect. \ref{sez2}, the basic motivations behind the use of Bardeen and Hayward metrics are reported. In Sect. \ref{sez3}, the Bardeen metric is critically analyzed, ending up with the corrections to entropy at late and early-times. The same is performed in Sect. \ref{sez4}, where the same is reported for the Hayward spacetime. The Page time is therefore studied in Sect. \ref{sez5}. Finally, in Sect. \ref{sez6}, we report our conclusions and perspectives.

\section{Islands, Hawking radiation and RBH}\label{sez2}

In view of BH Hawking radiation, limited within a region denoted as $R$, the density matrix of $R$ is typically determined by taking a partial trace over the complementary region $\overline R$. Adopting the recipe of  the minimal quantum extremal surface, as outlined in Refs.  \cite{Ryu:2006bv,Hubeny:2007xt,Engelhardt:2014gca}, certain systems lying on $\overline R$ are known as islands, $I (\subset \overline R)$. These should be excluded from the systems that are traced out, leading to the subsequent  entanglement entropy, in $R$, effectively determined by the systems in $R \cup I$.

Thus, the Hawking radiation entanglement entropy reads
\begin{equation}\label{formulagenerale}
 S(R)
 =
 \min \left\{\mathrm{ext}\left[
 \frac{\mathrm{Area}(\partial I)}{4G_{\rm N}}
 + S_{\rm matter}(R\cup I)
 \right]\right\} \ ,
\end{equation}
in which the prescription of the quantum extremal surface has been used\footnote{This formula allows for computing  entanglement entropy of Hawking radiation in $R$ since it employs the systems in $R \cup I$, thus not tracing away systems in the \emph{island regions}.}.

In the above relation,  $S_{\rm matter}(R\cup I)$ denotes the entanglement entropy of the matter fields in the region $R\cup I$, $\mathrm{Area}(\partial I)$ is the area of the extremal surface that forms the boundary of the region $I$, and $G_N$ is the Newton constant. The minimum is computed over $\mathcal I$, where $\partial I$ represents the boundary of the island, the quantum
extremal surface. Afterwards, the island rule was derived by using the \emph{replica method} for the gravitational path integral, and so as one applies the replica trick \cite{Callan:1994py,Holzhey:1994we,Calabrese:2009entanglement}
to gravitational theories,
one gets fixed boundaries due to the replica geometries. The replica trick allows for the same rule as described above to be derived, with the quantum extremal surface prescription playing a key role. This strengthens the concept of islands and suggests that the island conjecture may be applicable to all types of BHs, as supported by recent research \cite{Penington:2019kki,Hartman:2020swn,Goto:2020wnk}.

{

We investigate the effect of islands in the context of the aforementioned RBHs, supposing that the thermodynamics of these objects is equivalent to that of conventional BH. Motivated by this hypothesis, as RBHs exhibit horizons and asymptotic flatness, they clearly provide a suitable framework for exploring the island formula. }

Specifically, we focus on spherically symmetric, non-rotating metrics as the simplest approach for describing compact objects. Further, we investigate how the inclusion of regularity behavior at $r=0$ modifies the island formula compared to the Schwarzschild metric.

Hereafter we shall consider
\begin{equation}\label{abs}
ds^2 = -f(r)dt^2 + \frac{1}{f(r)}dr^2 + r^2(d\theta^2 + \sin^2\theta d\phi^2)\,,
\end{equation}
being this adaptable to our RBH solutions since $f(r)$ is assumed to be a smooth function. By singling it out, it is possible to define the kind of RBH under exam.

Below, we focus on Bardeen and Hayward solutions \cite{bardeen1968proceedings,Hayward:2006nonsingular}, motivating them since they provide topological charges and vacuum energy at $r=0$.

\section{Islands from the Bardeen metric}\label{sez3}

The Bardeen metric \cite{bardeen1968proceedings,Ayon-Beato:2000mjt} represents a solution of Einstein's field equations. It appears as non-rotating BH with \emph{topological charge}. The solution comes from Einstein-Maxwell equations describing  a magnetically-charged BH looking similar to the traditional Reissner-Nordstr\"{o}m BH solution, but without singularity at $r=0$.

Looking at Eq. \eqref{abs},  we thus write
\begin{equation}\label{blackening}
f (r) =  1 - \frac{2 M r^2}{(r^2 + q^2)^{3/2}}\,,
\end{equation}
where  $q$ and $M$ are the  charge and mass of the magnetic monopole, respectively. The limiting case, namely the Schwarzschild BH, is clearly recovered as $q\rightarrow{0}$.

This spacetime solution could be somehow reinterpreted as \emph{quasi-Kerr solution} \cite{Bambi:2014accretion}, since as it has been stated in Ref. \cite{Bambi:2013ufa}, the Bardeen metric is equivalent to the Kerr one only in its
rotating version\footnote{The rotation is obtained through the Newman-Janis algorithm \cite{NewmanJanis:1965algorithm,Newman:1965algorithmcharged}. By this property, several
rotating RBHs may be found through the Newman-Janis algorithm.}.

Following Ref. \cite{Ayon-Beato:2000mjt}, we can shift to \emph{Kruskal-like} coordinates\footnote{The procedure is the same performed in Kruskal coordinates. Since the BH is not the Schwarzschild one, we cannot claim that the coordinate change is exactly Kruskal, but rather a Kruskal replacement on a RBH solution, leading to the concept of Kruskal-like coordinates.}

The procedure is to work out  the tortoise coordinate

\begin{equation}\label{tortoise}
r_* = \int g_{rr}dr = \int \frac 1 {f(r)} dr\,,
\end{equation}
with the Finkelstein-like coordinates defined by the shifts:
\begin{align}
    u=t-r_*\,,\quad v=t+r_* \,,
\end{align}
allowing one to define the infinitesimal coordinates $
    du = dt -g_{rr}dr,     dv = dt + g_{rr}dr$, and the subsequent spacetime, rewritten by
\begin{equation}
    ds^2 = - f(r) dudv + r^2 d\Omega^2\,,
\end{equation}
where $d\Omega$ is the usual angular part of a given spherically symmetric spacetime, i.e., $d\Omega^2\equiv d\theta^2+\sin\theta^2d\phi^2$.

The Kruskal-like coordinates are \cite{Wang_2021}
\begin{align}
    U   =-e^{-\kappa_+t + \kappa_+r_*}\,,\quad    V   =e^{\kappa_+t + \kappa_+r_*}\,,
\end{align}
where $\kappa_+$ in the surface gravity calculated at the outer horizon. So to compute it, we get the $``+"$ root of $g_{tt} = 0$, which according to Ref. \cite{Mansoori_etall:2022} can be written as\footnote{The $``+"$ root implies that the corresponding radius is larger than the other root(s).}
\begin{equation}
\kappa_\pm = \frac{f'(r_\pm)}{2} \ .
\end{equation}

The metric can be finally recast under the useful form prompted by
\begin{equation}\label{kruskal_coord}
    ds^2 = -W^2(r){dUdV} + r^2 d\Omega^2\,,
\end{equation}
with
\begin{equation}
    W^2(r) = f(r) \frac{e^{-2\kappa_+r_*}}{\kappa_+^2}\,,
\end{equation}
where the \emph{weight function $W(r)$} represents the Jacobian to pass from one set of coordinates to another.

It is now necessary to work out horizons from the Bardeen spacetime. Given the analytic expression for the metric, the horizons results from
\begin{equation}\label{horizon_complete}
    1 - \frac{2Mr^2}{(r^2+q^2)^{3/2}} = 0\,,
\end{equation}
showing two positive distinct solutions\footnote{Any unphysical negative solution is clearly discarded into computation, consisting to non-relevant terms that do not contribute to the horizon computation.  }, say $r_-$ and $r_+$, where as usual we conventionally require $r_- < r_+$.

Motivated by the fact that topological charges are small as numerically found, see e.g. \cite{orl1}, Eq. \eqref{horizon_complete} can be easily solved under the prescription of small $q$. It is remarkable to note that we do not require large distances but small charges, i.e., we do not limit our treatment to large radii, but we expand around small $q$, implying $q/r \ll 1$, up to the third order in $q/r$, since $r>0$. Thus, we have
\begin{equation}
    f(r) \approx 1 -\frac {2M}r + \frac{3Mq^2}{r^3} = \frac{r^3 - 2Mr^2 + 3Mq^2}{r^3}\,.
\end{equation}

\noindent From Eq. \eqref{tortoise}, we obtain

\begin{equation}
r_* = r-\sum_{j}B_j\log(r-r_j)\,,
\end{equation}
having defined the auxiliary functions
\begin{equation}
B_j = \frac{M}{r_j}\frac{(2 r_j^2-3q^2)}{(4M - 3r_j)} \,, \,\,{\rm with}\quad j=1,2,3.
\end{equation}
where $r_j$'s are the roots of $r^3 - 2 Mr^2 + 3Mq^2=0$.
Explicitly we can write the tortoise coordinate as
\begin{align}\label{tortoise_approx}
    r_*  \approx r &- B_1\log(r-r_1) \\
    &- B_2\log(r-r_2)-B_3\log(r-r_3)\,.\nonumber
\end{align}

We have now all the ingredients to compute the entropy contribution without and with islands, as we report in the incoming subsection.

\subsection{Entropy without islands}\label{entropy_chapter}

The formula in Eq. \eqref{formulagenerale}  can be sorted out by means of the renormalizeed Newton's constant, $G_{N,r}$, under the form
\begin{equation}
\frac{1}{4G_{N,r}} \rightarrow \frac{ 1}{4G_N} + \frac{1}{\epsilon^2}\,,
\end{equation}
where $\epsilon$ is the cutoff scale used in the configuration that we intend to write up. In particular, following Refs. \cite{Wang_2021,Hashimoto_2020}, we invoke
\begin{subequations}
\begin{align}
    b_+ &= (t_b, b)\,,\\
    b_- &= (-t_b + i\frac \pi{\kappa_+}, b)\,,\\
    a_+ &= (t_a, a)\,,\\
    a_- &= (-t_a + i\frac \pi{\kappa_+}, a)\,,
\end{align}
\end{subequations}
where $b_\pm$ are the boundaries of the entanglement region $R$ and $a_\pm$ the boundaries of the island as shown in Fig. \ref{fig:penrose}.

\begin{widetext}

\begin{figure}[h]
\begin{center}
\begin{tikzpicture}[scale=1]

\node[coordinate] (center) at (0,0){};
\node[coordinate] (I) at (2,0) {I};
\node[coordinate] (II)   at (-2,0)   {II};
\node[coordinate] (III)  at (0, 2) {III};
\node[coordinate] (IV)   at (0,-2) {IV};
\node[coordinate] (Itop) at (2,2){Itop};
\node[coordinate] (IItop) at (-2,2){IItop};
\node[coordinate] (Ibot) at (2,-2){Ibot};
\node[coordinate] (IIbot) at (-2,-2){IIbot};

\node[shape=circle](I) at (0,1.3){$I$};
\node[shape=circle](R) at (2.4,0.8){$R$};

\node[coordinate] (IIleft) at (-4,0){IIleft};
\node[coordinate] (Iright) at (4,0){Iright};

\node[coordinate] (IIItop) at (0,3){IIItop};
\node[coordinate] (IVbot) at (0,-3){IVbot};

\node[shape=circle,fill=black, scale=0.3](a+) at (27:1.5)[label=below:$a_+$]{};
\node[shape=circle,fill=black, scale=0.3](b+) at (10:2.5)[label=below:$b_+$]{};

\node[shape=circle,fill=black, scale=0.3](a-) at (-27:-1.5)[label=below:$a_-$]{};
\node[shape=circle,fill=black, scale=0.3](b-) at (-10:-2.5)[label=below:$b_-$]{};

\draw[color=blue] (b+) .. controls (2.9,0.2) .. (Iright);
\draw[color=red] (a-) .. controls (0,1) .. (a+);

\draw[color=blue] (b-) .. controls (-2.9,0.2) .. (IIleft);

\draw (IIleft)--(IIItop)-- node[pos =0.2,above=0.2,sloped]{$r=r_-$}(Iright)--(IVbot)--cycle;
\draw (center) -- ++(-2,1.5) -- ++(0,3) --  (IIItop);
\draw (center) -- node[pos =0.7,above=0.2,sloped]{$r=r_+$} ++(2,1.5) -- node [right=0.2]{$r=0$} ++(0,3)[label=$r=0$] -- (IIItop);

\draw (center) -- ++(2,-1.5) -- ++(0,-3) -- (IVbot);
\draw (center) -- ++(-2,-1.5) -- ++(0,-3) -- (IVbot);

\end{tikzpicture}
\caption{Penrose diagram of the generic configuration in exam. The points $a$ and $b$ represent the boundaries of the island $I$ and the region $R$ respectively. The subscripts $\pm$ denotes the right and left wedge respectively.}
\label{fig:penrose}
\end{center}
\end{figure}

\end{widetext}

{Following Refs. \cite{Wang_2021,Hashimoto_2020}, we only employ the s-wave approximation for the Hawking radiation. In such a way, we also ignore gray body factors, leading, in the configuration \emph{without island}, to an entropy simply computed from quantum field theory recipe as}

\begin{equation}
    S_{\text{matter}} = \frac c3 \log{d(b_+, b_-)}\,,
\end{equation}
where $c$ is the central charge, as in \cite{Hashimoto_2020}, and $d$ is the distance.

The latter, in the Kruskal-like coordinates, Eq. \eqref{kruskal_coord}, can be written as
\begin{widetext}
\begin{equation}
\begin{split}\label{entropy_noisland}
        S_{\text{matter}} &= \frac{c}{6}\log \left[ W(b_-)W(b_+)(U(b_-)-U(b_+))(V(b_+)-V(b_-)) \right] \\
        &= \frac{c}{6}\log \left[ W^2(b)(e^{\kappa_+t_b + \kappa_+ r_*(b)} + e^{-\kappa_+t_b + \kappa_+ r_*(b)})(e^{\kappa_+t_b + \kappa_+ r_*(b)} + e^{\kappa_+t_b + \kappa_+ r_*(b)} \right]\\
        &= \frac c6 \log \left[ 4 W^2(b) e^{2\kappa_+ r_*(b))} \cosh^2({\kappa_+ t_b}) \right] \,.
\end{split}
\end{equation}
\end{widetext}
The corresponding behaviors at both late and early-times can be therefore analysed. Specifically, the behavior at late-times appears particularly interesting as the entropy could increase indefinitely eventually violating the Bekenstein bounds. Actually, for $t_b \gg 1$, Eq. \eqref{entropy_noisland} gives
\begin{equation}\label{entropy_noisland_late}
    S_{\text{matter}} \sim \frac{c}{3}\kappa_+ t_b\,,
\end{equation}
which, as above anticipated, diverges leaving the information paradox unsolved. As possible solution, we then include islands and check whether their inclusion would modify the corresponding entropy behavior.

\vspace{0.5cm}

\subsection{Entropy with islands}

{As stated earlier, we here include islands to check how they can modify the corresponding entropy behavior. Specifically, all multi-island configurations will be suppressed in the minimization procedure, if compared to the single-island configuration. Indeed, requiring that at early-times the entanglement entropy between radiation and RBH is small, the minimization of entropy might occur around $r=0$. Consequently, the first island contributes more significantly than other multi-islands configurations, that are far from the minimization region, $r=0$, see e.g. \cite{Hashimoto_2020,Wang_2021}. As multi-islands are suppressed by minimizing the entropy, we obtain }
\begin{widetext}
\begin{equation}
\begin{split}
    S_{\text{matter}} &=  \frac c3 \log{\left[\frac{d(a_+, a_-)d(b_+, b_-)d(a_+, b_+)d(a_-, b_-)}{d(a_+, b_-)d(a_-, b_+)}\right]}\\
    &= \frac c3 \log{d(b_+, b_-)d(a_+, a_-)}  \\
    &+\frac c3 \log \left[ \frac{(U(b_+)-U(a_+))(V(a_+)-V(b_+))(U(b_-)-U(a_-))(V(a_-)-V(b_-))}{(U(b_-)-U(a_+))(V(a_+)-V(b_-))(U(b_+)-U(a_-))(V(a_-)-V(b_+))} \right]\,.
\end{split}
\end{equation}
\end{widetext}

Using
\begin{equation}
    S_{\text{gen}} = \frac{2 \pi a^2}{G_N} + S_{\text{matter}}\,,
\end{equation}
in analogy to previous literature, see e.g. \cite{Wang_2021}, we get

\begin{widetext}
\begin{equation}\label{entropy_gen}
\begin{split}
       S_{\text{gen}}&= \frac{2 \pi a^2}{G_N} + \frac c6 \log \left[ 2^4 W^2(b)W^2(a) e^{2\kappa_+(r_*(a) + r_*(b))}\cosh^2({\kappa_+ t_b})\cosh^2({\kappa_+ t_a}) \right] \\
       &+\frac c3 \log \left[ \frac{\cosh{(\kappa_+ (r_*(a) - r_*(b)))} -\cosh{(\kappa_+ (t_a - t_b))}}{\cosh{(\kappa_+ (r_*(a) - r_*(b)))} + \cosh{(\kappa_+ (t_a + t_b))}}\right]\,.
\end{split}
\end{equation}
\end{widetext}

{The early-time behavior can be compared with late-times. In the former case, we assume $t_a, t_b \ll r_+$. Further, we note that if at least one island exists, its place  should be close to $r = 0$ since, at early-times, the entanglement entropy  between the radiation and the RBH is small and the extremal surface that can minimize the entropy has to lie very close to $r = 0$, or do not exist at all}.

Bearing this in mind, one has
\begin{subequations}
\begin{align}
    &\cosh{(\kappa_+ (r_*(a) - r_*(b)))} \gg \cosh{(\kappa_+ (t_a - t_b))}\,,\\
    &\cosh{(\kappa_+ (r_*(a) - r_*(b)))} \gg \cosh{(\kappa_+ (t_a + t_b))}\,,
\end{align}
\end{subequations}
and rewriting Eq. \eqref{entropy_gen} as
\begin{widetext}
\begin{align}
     S_{\text{gen}}&\approx \frac{2 \pi a^2}{G_N} +\frac c6 \log \left[ 2^4 W^2(b)W^2(a) e^{2\kappa_+(r_*(a) + r_*(b))}\cosh^2({\kappa_+ t_b})\cosh^2({\kappa_+ t_a}) \right]\nonumber\\
     &\approx \frac{2 \pi a^2}{G_N} +\frac c6 \log [W^2(a)e^{2\kappa_+r_*(a)}] + \text{functions of $b$ only}\nonumber\\
     &= \frac{2 \pi a^2}{G_N} +\frac c6 \log [f(a)] + \text{functions of $b$ only.}
\end{align}
\end{widetext}
we can now extremize $S_{\text{gen}}$ with respect to $a$. This yields
$\frac{4 \pi a}{G_N} + \frac c6 \frac{f'(a)}{f(a)}=0$ or, more explicitly
\begin{align}
    \frac{4 \pi a}{G_N} + \frac c6 \frac{2 M \left(a^3-2 a q^2\right)}{\left(a^2+q^2\right)
   \left(\left(a^2+q^2\right)^{3/2}-2 a^2 M\right)}=0\,.
\end{align}
Expanding the l.h.s up to the third order in $a$, we get
\begin{equation}
    a \approx \sqrt{\frac{q^3}{cMG_N}\frac{6 \pi q^3 - c M G_N}{2M-3q}}\,.
\end{equation}
{Therefore, in principle, the island cannot form as the the Newton's constant contribution dominates over the third order topological term, requiring
\begin{equation}
q<\left(\frac{cMG_N}{6 \pi}\right)^{\frac{1}{3}}\,.
\end{equation}
However, this is not a
physical bound, but just a bound due to the corresponding early time approximation. Indeed, since usually the early-time entanglement entropy evolution is dominated by non-island saddle, the existence of islands does not seem necessary
in this regime. For a further discussion about the plausible values of $q$, allowed in the case of Bardeen RBH, with $c\geq0$, we refer to  Ref. \cite{orl1}.}

The same procedure can be carried out studying late-times. Here we assume $t_a,t_b \gg b$ which allow us to make the following approximation, see also \cite{Mansoori_etall:2022},

\begin{equation}
    d(a_+,a_-) \approx d(b_+,b_-)\approx d(a_\pm, b_\mp) \ll d(a_\pm, b_\pm)\,,
\end{equation}
which leads to
\begin{equation}
    S_{\text{matter}} \approx \frac{c}{3} \log{[d(a_+, b_+)d(a_-, b_-)]}\,.
\end{equation}
After some algebra this can be rewritten as
\begin{widetext}
\begin{equation}
    S_{\text{matter}}=\frac c3 \log{\left| \frac{1}{\kappa_+} \sqrt{f(a)f(b)}e^{-2\kappa_+(a_* + b_*)} \left[e^{2\kappa_+ a_*} + e^{2\kappa_+ b_*} - 2e^{\kappa_+(a_* + b_*)}\cosh{(t_a-t_b)}\right] \right|}\,,
\end{equation}
\end{widetext}
\noindent where $a_* = r_*(a)$ and $b_* = r_*(b)$ have been defined.

Maximizing with respect to $t_a$, we can get
\begin{equation}
\begin{split}
    \max_{t_a}S_{\text{matter}}&=\frac{2c}{3}\log{\left| \frac{e^{\kappa_+a_*}-e^{\kappa_+b_*}}{\kappa_+}\right|}\\
    &+ \frac{c}{6} \log{\left| f(a)f(b) e^{-2\kappa_+(a_* + b_*)} \right| }\,,
\end{split}
\end{equation}
where $t_a = t_b \equiv t$ and so, in this case, we need to find out the island position, by maximizing $S_{\text{gen}}$ with respect to $a$.

Hence,  we set
$    \partial_a S_{\text{gen}} = \partial_a \left( \frac{2\pi a^2}{G_N} + \max_{t_a} S_{\text{matter}} \right) = 0$,
resulting in
\begin{equation}\label{aeq}
\frac{4\pi a}{G_N} + \frac c6 \left[ \frac{f'(a) - 2\kappa_+}{f(a)} \right] + \frac{2c}{3} \frac{\kappa_+ e^{\kappa_+ a_*}}{f(a)(e^{\kappa_+a_*} + e^{\kappa_+b_*})} = 0\,.
\end{equation}
We now assume that the island is located very close, but outside the horizon, which means $a \approx r_+$ and $a > r_+$. This allows us to expand $f(a)$ around $r_+$ as
\begin{align}\label{approx_f}
f(a) \approx f(r_+) + f'(r_+)(a-r_+) = 2\kappa_+ r_+ \frac{a-r_+}{r_+}\,,
\end{align}
since $f(r_+) = 0$ and $f'(a)\approx f'(r_+) = 2\kappa_+$. From Eq. \eqref{tortoise} we get
\begin{equation}\label{approx_a_star}
a_* \approx \int^a \frac{1}{2\kappa_+ (r-r_+)}dr = \frac{1}{2\kappa_+} \log{\frac{a-r_+}{r_+}}\,,
\end{equation}
using $dr = r_+ d\left(\frac{r - r_+}{r_+}\right)$.

Consequently, the equation for $a$, Eq. \eqref{aeq},  acquires the form
\begin{equation}
    \frac{4\pi r_+}{G_N} + \frac{c}{3}\frac{\sqrt{\frac{a-r_+}{r_+}}}{r_+\frac{a-r_+}{r_+}\left(\sqrt{\frac{a-r_+}{r_+}} + e^{\kappa_+b_*}\right)}\approx0\,,
\end{equation}
and so, keeping the leading term in $(a-r_+)$ only, we write

\begin{equation}
    \frac{4\pi r_+}{G_N} + \frac{c}{3}\frac{e^{-\kappa_+b_*}}{\sqrt{r_+}\sqrt{a-r_+}}\approx0\,,
\end{equation}
whose solution, provides the island position as
\begin{equation}
    a \approx r_+ + \left( \frac {G_N c}{12 \pi r_+} \right)^2\frac 1{r_+} e^{-2\kappa_+ b_*}\,,
\end{equation}
that manifestly depends on $b$ only.

We can then plug the tortoise coordinates, Eq. \eqref{tortoise_approx}, into $e^{-2\kappa_+ b_*}$,  giving
\begin{widetext}
\begin{equation}
    a \approx r_+ + \left( \frac {G_N c}{12 \pi r_+} \right)^2 \frac 1{r_+} e^{-2\kappa_+ b} (b-r_1)^{2\kappa_+ B_1}(b-r_2)^{2\kappa_+ B_2}(b-r_3)^{2\kappa_+ B_3}\,,
\end{equation}
\end{widetext}
and then we end up with inserting the above value for $a$ in the expression for the generalized entropy, Eq. \eqref{entropy_gen}.

Recalling that we are studying the late time behavior of the solutions, we consider the following approximations
\begin{subequations}
\begin{align}
    \cosh{(\kappa_+ (t_a +t_b))} &\approx 2 \cosh{(\kappa_+ t_a)} \cosh{(\kappa_+ t_b)}\,,\\
    \cosh{(\kappa_+ (t_a + t_b))}&\gg \cosh{(\kappa_+ (a_* - b_*))}
\end{align}
\end{subequations}
and, using the definition for $t$ with the additional requirement $t \gg b \gg r_+$, we can rewrite Eq. \eqref{entropy_gen} as

\begin{widetext}
\begin{align}\label{entropy_gen_intermedio}
S_{\text{gen}}      &=\frac{2\pi a^2}{G_N} + \frac c6\log{[W^2(b)W^2(a)]}
       +\frac c3 \log \left| \frac{1 -2e^{-\kappa_+ (b_* - a_*)}}{1 + e^{\kappa_+(a_* - b_* -2t)}}\right| + \frac {2c}3 \kappa_+ b_*\\
       &=\frac{2\pi a^2}{G_N} + \frac c6\log{[W^2(b)W^2(a)]} +\frac c3 \log \left| -2e^{-\kappa_+ (b_* - a_*)} - e^{\kappa_+(a_* - b_* -2t)}\right| + \frac {2c}3 \kappa_+ b_*\,,\nonumber
\end{align}
\end{widetext}

\noindent where we computed the  logarithm expansion and used $
    \cosh{(\kappa_+ (a_* - b_*))} \approx \frac 12 e^{\kappa_+ (a_* - b_*)}$.

We note that the only time dependent term in Eq. \eqref{entropy_gen_intermedio} decreases exponentially, i.e., at late-times $S_{\text{gen}}$ becomes constant. The corresponding value reads
\begin{equation}
\begin{split}
    S_{\text{gen}}&= \frac{2\pi a^2}{G_N} + \frac c6\log{[W^2(b)W^2(a)]} +\\
       &+\frac c3 \log \left| -2e^{-\kappa_+ (b_* - a_*)} - 1\right| + \frac {2c}3 \kappa_+ b_*\,,
\end{split}
\end{equation}
representing the asymptotic value of the entropy entering the Hawking entropy.

Finally, for $\frac{a - r_+}{r_+} = \epsilon \ll 1$ we have
\begin{equation}
W^2(a) = \frac{f(a)}{\kappa_+^2} e^{-2\kappa_+ a_*}\,,
\end{equation}

that, taking into account Eqs. \eqref{approx_f} and \eqref{approx_a_star}, can be approximated by

\begin{equation}
W^2(a) \approx \frac{2 r_+ \kappa_+}{\kappa_+^2}\epsilon \exp{\left[- 2 \kappa_+ \frac{1}{2 \kappa_+} \log \epsilon\right]}= \frac{2 r_+}{\kappa_+} \,.
\end{equation}
At leading order in $\epsilon$, we have
\begin{equation}\label{entropy_final_generic}
\begin{split}
    S_{\text{gen}} &\approx \frac{2\pi r_+^2}{G_N} + \frac c6 \log{[\frac{2 r_+}{\kappa_+}W^2(b)]} + \frac{2c}{3}\kappa_+ b_*\\
    &=\frac{2\pi r_+^2}{G_N} + \frac c6 \log{\left[\frac{2 r_+}{\kappa_+^3}f(b) e^{2\kappa_+b_*}\right]}\,,
\end{split}
\end{equation}
where, using Eqs. \eqref{blackening} and  \eqref{tortoise_approx}, we can get
\begin{widetext}
\begin{multline}\label{entropy_final_bardeen}
    S_{\text{gen}} \approx \frac{2\pi r_+^2}{G_N} + \frac c6 \log{\left[\frac{2 r_+}{\kappa_+}\left( 1 - \frac{2M b^2}{(b^2 + q^2)^{3/2}} \right)\right]} +\\
    + \frac{2c}{3}\kappa_+ \left[ b - \frac{M(-3q^2 + 2 r_1^2)}{(4M - 3r_1)r_1}\log{(b-r_1)} - \frac{M(-3q^2 + 2 r_2^2)}{(4M - 3r_2)r_2}\log{(b-r_2)} - \frac{M(-3q^2 + 2 r_3^2)}{(4M - 3r_3)r_3}\log{(b-r_3)}\right]\,.
\end{multline}
\end{widetext}
The aforementioned expression refers to the total entropy composed by the standard Hawking part summed with the island correction. The behavior of our finding is compared with the standard Schwarzschild case \cite{Hashimoto_2020} in Fig. \ref{fig:entropy}, where $S_{\text{Schwarz.}}$ comes from Eq. \eqref{entropy_final_generic} using
\begin{equation}\label{schwarzschild_functions}
    \begin{split}
        &f(r) = 1 - \frac{2M}{r}\\
        &r_* = r - 2M + 2M \log(r- 2M) \, .
    \end{split}
\end{equation}

\section{The Hayward spacetime}\label{sez4}

Recently, there has been a need to extend the Schwarzschild de-Sitter solution by including a vacuum energy term in a regular configuration, which led to the introduction of the Hayward solution \cite{Hayward:2006nonsingular}. This regular configuration is similar in physical interpretation to the Bardeen solution, with a central flatness. The corresponding lapse function for the regular black hole implies a specific matter energy-momentum tensor that is de Sitter at the core and recovers Minkowski at asymptotic distances. The extra-parameter responsible for flatness, denoted by $\Lambda$, can be identified with a magnetic charge for a given non-linear electrodynamic theory, making the Hayward metric a solution of such classes of theories as well as the Bardeen metric \cite{Malafarina:2022wmx}. However, the Hayward spacetime exhibits some inconsistencies at the level of higher order curvature invariants, as discussed in \cite{Zhou:2023analyticRBH,Giacchini:2021highrOrderDerivative}. Therefore, we wonder whether it would be possible to observe these characteristics at the level of islands.

In the Hayward RBH the function $f(r)$ in Eq. \eqref{abs} takes the form
\begin{equation}
f (r) =  1 - \frac{2 M r^2}{r^3 + 2 M \Lambda^2}\,,
\end{equation}
where, again $M$, is the point-like mass of the RBH, whereas $\Lambda$ is intimately related to the constant term mimicking vacuum energy, $\Lambda_{vac}$. In fact, as  $r$ is small enough, $f(r)\simeq 1-r^2/\Lambda^2=1-\Lambda_{vac}^2 r^2$, miming a de Sitter phase. For our purposes, once fixed $M$, we focus on small and large values of $\Lambda$, corresponding to large and small vacuum energy contributions. In our findings, we  take the sign of $f(r)$ to be always positive, guaranteeing that the island position is always placed over $r_+$.

In analogy to the Bardeen case, we perform below the same computation to get the Hawking entropy corrections without and with islands.

\subsection{Islands from the Hayward spacetime}

As the derivation in Sec. \ref{sez3} is valid for every spherically symmetric and static metric, we can immediately write
\begin{equation}
    a \approx r_+ + \left( \frac {G_N c}{12 \pi r_+} \right)^2\frac 1{r_+} e^{-2\kappa_+ b_*}\,,
\end{equation}
and
\begin{equation}
    S_{\text{gen}} \approx \frac{2\pi r_+^2}{G_N} + \frac c6 \log{\left[\frac{2 r_+}{\kappa_+^3}f(b) e^{2\kappa_+b_*}\right]}\,.
\end{equation}
In this case however, the tortoise coordinate can be exactly computed to give
\begin{align}\label{tortoise_hayward}
    r_* =r - 2M\sum_{j=1}^{3} H_j \log (r-r_j) \,,
\end{align}
having defined
\begin{equation}
    H_j = \frac{r_j}{4M - 3r_j}, \, \text{with } j=1,2,3 \, .
\end{equation}
Here the $r_j$'s denote the positive roots of the polynomial $2 M\Lambda^2-2 M r^2+r^3$.
In order to ensure the presence of at least one event horizon we need to impose $\Lambda \leq \frac{4}{3\sqrt{3}} M$, because otherwise the discriminant of the polynomial is negative resulting in one negative root and two complex roots.

We can use Eq. \eqref{tortoise_hayward} to obtain
\begin{equation}
    a \approx r_+ + \left( \frac {G_N c}{12 \pi r_+} \right)^2\frac 1{r_+} \left(b-r_1\right)^{H_1}\left(b-r_2\right)^{H_2}\left(b-r_3\right)^{H_3}\,,
\end{equation}
and
\begin{widetext}
\begin{equation}\label{entropy_final_hayward}
\begin{split}
    &S_{\text{gen}} \approx \frac{2\pi r_+^2}{G_N} + \frac c6 \log{\left[\frac{2 r_+}{\kappa_+^3}\left( 1-\frac{2Mb^2}{b^3 + 2M \Lambda^2} \right) e^{2\kappa_+b_*}\right]}\\
    & + \frac{2c\kappa_+}{6} \left[b - 2M\left( \frac{r_1 \log{(b - r_1)}}{4M - 3r_1} + \frac{r_2 \log{(b - r_2)}}{4M - 3r_2} + \frac{r_2 \log{(b - r_3)}}{4M - 3r_3} \right)\right]\,.
\end{split}
\end{equation}
\end{widetext}

\begin{figure}
    \centering
    \includegraphics[scale=.9]{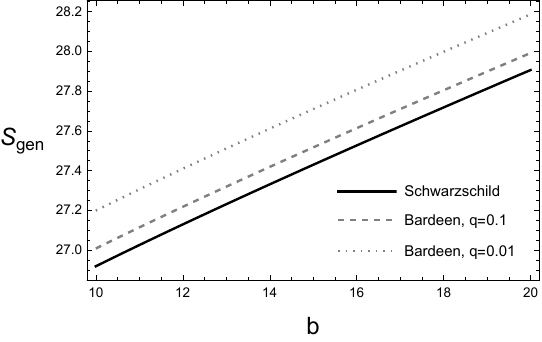}
    \includegraphics[scale=.9]{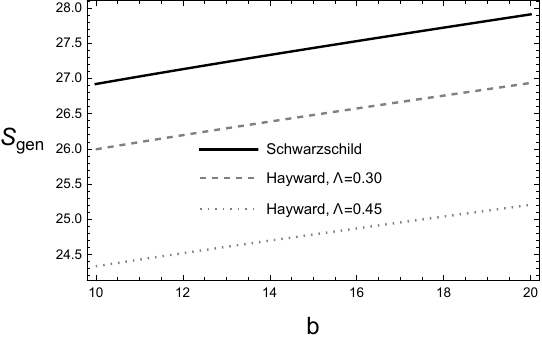}
    \caption{Behaviors of generalized entropy for the Bardeen and Hayward spacetimes generated from Eq. \eqref{entropy_final_bardeen} (Bardeen) and Eq. \eqref{entropy_final_hayward} (Hayward) choosing $M = G = c = 1$. The Schwarzschild entropy as been calculated using Eq. \eqref{entropy_final_generic}. }
    \label{fig:entropy}
\end{figure}

Formally both solutions, Eq. \eqref{entropy_final_bardeen} and Eq. \eqref{entropy_final_hayward}, appear similar. Moreover, the result above found appears compatible with previous findings, see e.g. \cite{Kim:2021entanglement}. However, the physical mechanisms behind the aforementioned results is extremely different, leading to corrections that look different as well. The above computed entropy clearly deviates from the Bardeen case. Analogy and differences are prompted in Fig. \ref{fig:entropy}.

\section{The Page time}\label{sez5}

We previously studied the entropy behavior of our system both at early and late-times. We saw that initially it increases linearly in time (see Eq. \eqref{entropy_noisland_late}) but, when enough radiation is emitted, an island is formed and the entropy remains constant as in Eq. \eqref{entropy_final_bardeen}. The time at which this transition happens is called \emph{Page time}. More generally the Page time is defined as the time when the entropy becomes constant and it is usually indicated by $t_{\text{Page}}$. Insights on how the Page time can be related to other quantities like the \emph{Scrambling time} or the \emph{Hartman-Maldacena time} can be found in Refs. \cite{Saha:2021mutualInfIsland,RoyChowdhury:2022mutualInfPage,RoyChowdhury:2023mutualSchdeSit}. We estimated $t_{\text{Page}}$ imposing that the \emph{entropy in the configuration without island is equal to that with island}.

We thus perform this computation, starting from Eqs. \eqref{entropy_noisland_late} and  \eqref{entropy_final_bardeen} and imposing
\begin{equation}
    \frac c3 \kappa_+ t_{\text{Page}} \approx \frac{2 \pi r_+^2}{G_N}\,,
\end{equation}
obtaining a Page time of the form
\begin{equation}\label{page_time}
    t_{\text{Page}} \approx \frac{6 \pi r_+^2}{cG_N \kappa_+ }\,,
\end{equation}
where we only kept the Bekenstein-Hawking term, i.e. $\frac{2 \pi r_+^2}{G_N}$, for the sake of simplicity.

Recalling the BH temperature
\begin{equation}
T_{\text{BH}} = \frac{\kappa_+}{2\pi}\,,
\end{equation}
and assuming that the regular configuration provides the same temperature, as motivated by recent studies, see e.g. \cite{IlichKruglov:2021pdw}, we take $T_{\text{BH}}=T_{\text{RBH}}$, with the latter the corresponding effective temperature for RBHs.

In Tab. \ref{tab:page1}, we report some numerical results for the Page time in our spacetime configurations. Particular attention has been devoted to the relative variation with respect to the Schwarzschild spacetime, i.e.,
\begin{equation}
\delta t\equiv \frac{\Delta t}{t_S}=\frac{t_{B;H}-t_S}{t_S} \, .
\end{equation}

\begin{table}[h!]
\centering
\begin{tabular}{c|c c c c}
\hline
\hline
    \text{} & \text{{\bf Schwarzschild}} & \text{{\bf Bardeen}} & \text{{\bf Hayward}} \\
    \hline
    $\,$ & $\quad$ & $q=0.005$ & $\Lambda=0.15$ & $\,$ \\
\hline
    $t_{\text{Page}}$ & 150.80 & 150.80 & 151.67 \\
    $\delta t$ & 0 & 1.1719$\times 10^{-10}$ & 0.0058213 \\
\hline
\hline
    $\,$ & $\quad$ &   $q = 0.01$ & $\Lambda = 0.30$ & $\,$\\
\hline
 $t_{\text{Page}}$ & 150.80 & 150.80 & 154.72 \\
 $\delta t$ & 0 & 1.8753$\times 10^{-9}$ & 0.025999 \\
\hline
\hline
    $\,$ & $\quad$ &   $q = 0.015$ & $\Lambda = 0.45$ & $\,$\\
\hline
 $t_{\text{Page}}$ & 150.80 & 150.80 & 161.73 \\
 $\delta t$ & 0 & 9.4957$\times 10^{-9}$ & 0.072549 \\
\hline
\end{tabular}
\caption{Table of indicative values generated from Eq. \eqref{page_time} with the above choices of free parameters and $M = G = c = 1 $. The most significant departures from the Schwarzschild case are found for Hayward spacetime, while in the Bardeen configuration, the outcomes appear to be negligibly small. }
\label{tab:page1}
\end{table}

As emphasized above, the presented numerical results show that there is no clear evidence in favor of discrepancies occurring for RBHs than BHs. This means that the thermodynamics, as well as islands, are comparable to standard solutions, see e.g. \cite{Wald:1999vt}. More relevant departures are found in the context of Hayward solution, whereas the Bardeen spacetime does not show significant changes compared with the Schwarzschild solution. This implies that the presence of vacuum energy regular core in the Hayward solution provides more significant changes in our findings. This can be explained by the fact that topological charges are expected to have a weak contribution to the energy-momentum tensor of a given solution. On the other hand, the strength of vacuum energy is not limited, and can be either small or large depending on the specific $\Lambda_{vac}$ value in the metric, which consequently results in more noticeable modifications.

\section{Final remarks and outlook}\label{sez6}

We studied how the presence of non-singular regions for RBH configurations influences the island formula. To do so, we worked out the Bardeen and Hayward spacetimes and evaluated the corresponding thermodynamics of such objects, computing the island formula for both these metrics.

We distinguished two main cases associated with late and early-times, where we approximated the corresponding islands. Thus, we emphasized in which regions the islands can exist, underlying how the effects of topological charge and vacuum energy influence our findings. We pointed out our findings in the regime of arbitrary radii, involving small topological charges and small/large vacuum energy magnitudes. As those configurations can frame out compact objects, we compared our outcomes with respect to the Schwarzschild BH, and, consequently, we examined the effects of regularity  at $r=0$ and checked the main changes expected as $q$ and $\Lambda$ vanish.

We also computed the Page time, which represents the point at which the entanglement entropy of the black hole is half-saturated, and compared it with previous findings. Our results showed that the effects of the regular configurations we investigated were more pronounced when the free parameters of our RBHs were tuned, particularly in the case of the Hayward metric. However, the Bardeen spacetime did not deviate significantly from the Schwarzschild BH. Therefore, while it was evident that the presence of non-singular behavior modified the entropy contribution to islands, the predicted changes induced by regular solutions did not significantly alter the results obtained for the Schwarzschild BH. {In general, the  increased discrepancy of the Page time between the Schwarzschild
and Hayward spacetimes can be attributed to the presence of non-zero vacuum energy contribution that arises in the Hayward spacetime. Indeed, the latter metric corresponds to a spacetime that exhibits a non-zero cosmological constant-like contribution in the regions where islands appear. In this respect, it has been shown that the gravitational charge inducing the Hayward metric is not exactly the gravitational mass as it happens for a standard BH \cite{Luongo:2023aib}. This may suggest that the presence of $\Lambda$ determines a significant departure from the genuine Schwarzschild solution as its magnitude increases.}

In view of our results, it appears useful to investigate the island formula in other RBHs, including the effects of rotation and/or working out non linear electrodynamic contributions to the lapse function. Moreover, the task of using regular solution with compact object is still debated, so the need of working out rotating solutions and/or those providing quadrupole terms would appear interesting for future works. As future development we will focus  on  metrics, exhibiting horizons, that however may also show unphysical island regions and/or metrics that exhibit more than one island configurations.

\begin{acknowledgments}
OL is grateful to Carlo Cafaro, Roberto Giamb\`o, Daniele Malafarina and Hernando Quevedo for fruitful discussions on the subject of this paper. SM acknowledges financial support from “PNRR MUR project PE0000023-NQSTI”.
\end{acknowledgments}


\begin{thebibliography}{77}
\expandafter\ifx\csname natexlab\endcsname\relax\def\natexlab#1{#1}\fi
\expandafter\ifx\csname bibnamefont\endcsname\relax
  \def\bibnamefont#1{#1}\fi
\expandafter\ifx\csname bibfnamefont\endcsname\relax
  \def\bibfnamefont#1{#1}\fi
\expandafter\ifx\csname citenamefont\endcsname\relax
  \def\citenamefont#1{#1}\fi
\expandafter\ifx\csname url\endcsname\relax
  \def\url#1{\texttt{#1}}\fi
\expandafter\ifx\csname urlprefix\endcsname\relax\def\urlprefix{URL }\fi
\providecommand{\bibinfo}[2]{#2}
\providecommand{\eprint}[2][]{\url{#2}}

\bibitem[{\citenamefont{Abbott et~al.}(2016)}]{LIGOScientific:2016aoc}
\bibinfo{author}{\bibfnamefont{B.~P.} \bibnamefont{Abbott}}
  \bibnamefont{et~al.} (\bibinfo{collaboration}{LIGO Scientific, Virgo}),
  \bibinfo{journal}{Phys. Rev. Lett.} \textbf{\bibinfo{volume}{116}},
  \bibinfo{pages}{061102} (\bibinfo{year}{2016}), \eprint{1602.03837}.

\bibitem[{\citenamefont{Akiyama et~al.}(2019)}]{EventHorizonTelescope:2019dse}
\bibinfo{author}{\bibfnamefont{K.}~\bibnamefont{Akiyama}} \bibnamefont{et~al.}
  (\bibinfo{collaboration}{Event Horizon Telescope}),
  \bibinfo{journal}{Astrophys. J. Lett.} \textbf{\bibinfo{volume}{875}},
  \bibinfo{pages}{L1} (\bibinfo{year}{2019}), \eprint{1906.11238}.

\bibitem[{\citenamefont{{Volonteri} et~al.}(2021)\citenamefont{{Volonteri},
  {Habouzit}, and {Colpi}}}]{2021NatRP...3..732V}
\bibinfo{author}{\bibfnamefont{M.}~\bibnamefont{{Volonteri}}},
  \bibinfo{author}{\bibfnamefont{M.}~\bibnamefont{{Habouzit}}},
  \bibnamefont{and} \bibinfo{author}{\bibfnamefont{M.}~\bibnamefont{{Colpi}}},
  \bibinfo{journal}{Nature Reviews Physics} \textbf{\bibinfo{volume}{3}},
  \bibinfo{pages}{732} (\bibinfo{year}{2021}), \eprint{2110.10175}.

\bibitem[{\citenamefont{Penrose}(1965)}]{Penrose:65collapse}
\bibinfo{author}{\bibfnamefont{R.}~\bibnamefont{Penrose}},
  \bibinfo{journal}{Phys. Rev. Lett.} \textbf{\bibinfo{volume}{14}},
  \bibinfo{pages}{57} (\bibinfo{year}{1965}).

\bibitem[{\citenamefont{Hawking et~al.}(1970)\citenamefont{Hawking, Penrose,
  and Bondi}}]{Hawking:70sing}
\bibinfo{author}{\bibfnamefont{S.~W.} \bibnamefont{Hawking}},
  \bibinfo{author}{\bibfnamefont{R.}~\bibnamefont{Penrose}}, \bibnamefont{and}
  \bibinfo{author}{\bibfnamefont{H.}~\bibnamefont{Bondi}},
  \bibinfo{journal}{Proceedings of the Royal Society of London. A. Mathematical
  and Physical Sciences} \textbf{\bibinfo{volume}{314}}, \bibinfo{pages}{529}
  (\bibinfo{year}{1970}).

\bibitem[{\citenamefont{Hawking}(1976)}]{Hawking:76breakdown}
\bibinfo{author}{\bibfnamefont{S.~W.} \bibnamefont{Hawking}},
  \bibinfo{journal}{Phys. Rev. D} \textbf{\bibinfo{volume}{14}},
  \bibinfo{pages}{2460} (\bibinfo{year}{1976}).

\bibitem[{\citenamefont{Hawking}(1975)}]{Hawking:75particle}
\bibinfo{author}{\bibfnamefont{S.~W.} \bibnamefont{Hawking}},
  \bibinfo{journal}{Communications in Mathematical Physics}
  \textbf{\bibinfo{volume}{43}}, \bibinfo{pages}{199} (\bibinfo{year}{1975}).

\bibitem[{\citenamefont{Page}(1993)}]{Page:93information}
\bibinfo{author}{\bibfnamefont{D.~N.} \bibnamefont{Page}},
  \bibinfo{journal}{Physical Review Letters} \textbf{\bibinfo{volume}{71}},
  \bibinfo{pages}{3743} (\bibinfo{year}{1993}).

\bibitem[{\citenamefont{Page}(2013)}]{Page:2012time}
\bibinfo{author}{\bibfnamefont{D.~N.} \bibnamefont{Page}},
  \bibinfo{journal}{Journal of Cosmology and Astroparticle Physics}
  \textbf{\bibinfo{volume}{2013}}, \bibinfo{pages}{028} (\bibinfo{year}{2013}).

\bibitem[{\citenamefont{Penington}(2020)}]{penington:2020entanglement}
\bibinfo{author}{\bibfnamefont{G.}~\bibnamefont{Penington}}
  (\bibinfo{year}{2020}), \eprint{1905.08255}.

\bibitem[{\citenamefont{Almheiri
  et~al.}(2019{\natexlab{a}})\citenamefont{Almheiri, Engelhardt, Marolf, and
  Maxfield}}]{Almheiri:2019bulk}
\bibinfo{author}{\bibfnamefont{A.}~\bibnamefont{Almheiri}},
  \bibinfo{author}{\bibfnamefont{N.}~\bibnamefont{Engelhardt}},
  \bibinfo{author}{\bibfnamefont{D.}~\bibnamefont{Marolf}}, \bibnamefont{and}
  \bibinfo{author}{\bibfnamefont{H.}~\bibnamefont{Maxfield}},
  \bibinfo{journal}{Journal of High Energy Physics}
  \textbf{\bibinfo{volume}{2019}} (\bibinfo{year}{2019}{\natexlab{a}}).

\bibitem[{\citenamefont{Almheiri
  et~al.}(2020{\natexlab{a}})\citenamefont{Almheiri, Mahajan, Maldacena, and
  Zhao}}]{Almheiri:2020semiclassical}
\bibinfo{author}{\bibfnamefont{A.}~\bibnamefont{Almheiri}},
  \bibinfo{author}{\bibfnamefont{R.}~\bibnamefont{Mahajan}},
  \bibinfo{author}{\bibfnamefont{J.}~\bibnamefont{Maldacena}},
  \bibnamefont{and} \bibinfo{author}{\bibfnamefont{Y.}~\bibnamefont{Zhao}},
  \bibinfo{journal}{Journal of High Energy Physics}
  \textbf{\bibinfo{volume}{2020}} (\bibinfo{year}{2020}{\natexlab{a}}).

\bibitem[{\citenamefont{Almheiri
  et~al.}(2019{\natexlab{b}})\citenamefont{Almheiri, Mahajan, and
  Maldacena}}]{almheiri:2019islands}
\bibinfo{author}{\bibfnamefont{A.}~\bibnamefont{Almheiri}},
  \bibinfo{author}{\bibfnamefont{R.}~\bibnamefont{Mahajan}}, \bibnamefont{and}
  \bibinfo{author}{\bibfnamefont{J.}~\bibnamefont{Maldacena}}
  (\bibinfo{year}{2019}{\natexlab{b}}), \eprint{1910.11077}.

\bibitem[{\citenamefont{Omidi}(2022)}]{Omidi:2022hyperscaling}
\bibinfo{author}{\bibfnamefont{F.}~\bibnamefont{Omidi}},
  \bibinfo{journal}{Journal of High Energy Physics}
  \textbf{\bibinfo{volume}{2022}} (\bibinfo{year}{2022}).

\bibitem[{\citenamefont{Yu and Ge}(2021)}]{yu2021pageDilaton}
\bibinfo{author}{\bibfnamefont{M.-H.} \bibnamefont{Yu}} \bibnamefont{and}
  \bibinfo{author}{\bibfnamefont{X.-H.} \bibnamefont{Ge}},
  \emph{\bibinfo{title}{Page curves and islands in charged dilaton black
  holes}} (\bibinfo{year}{2021}), \eprint{2107.03031}.

\bibitem[{\citenamefont{Yu et~al.}(2022)\citenamefont{Yu, Lu, Ge, and
  Sin}}]{Yu:2022islandSuperradiance}
\bibinfo{author}{\bibfnamefont{M.-H.} \bibnamefont{Yu}},
  \bibinfo{author}{\bibfnamefont{C.-Y.} \bibnamefont{Lu}},
  \bibinfo{author}{\bibfnamefont{X.-H.} \bibnamefont{Ge}}, \bibnamefont{and}
  \bibinfo{author}{\bibfnamefont{S.-J.} \bibnamefont{Sin}},
  \bibinfo{journal}{Physical Review D} \textbf{\bibinfo{volume}{105}}
  (\bibinfo{year}{2022}).

\bibitem[{\citenamefont{Yu and Ge}(2023)}]{Yu:2023islandGenDilaton}
\bibinfo{author}{\bibfnamefont{M.-H.} \bibnamefont{Yu}} \bibnamefont{and}
  \bibinfo{author}{\bibfnamefont{X.-H.} \bibnamefont{Ge}},
  \bibinfo{journal}{Physical Review D} \textbf{\bibinfo{volume}{107}}
  (\bibinfo{year}{2023}).

\bibitem[{\citenamefont{Geng and Karch}(2020)}]{Geng:2020massiveIsland}
\bibinfo{author}{\bibfnamefont{H.}~\bibnamefont{Geng}} \bibnamefont{and}
  \bibinfo{author}{\bibfnamefont{A.}~\bibnamefont{Karch}},
  \bibinfo{journal}{JHEP} \textbf{\bibinfo{volume}{09}}, \bibinfo{pages}{121}
  (\bibinfo{year}{2020}), \eprint{2006.02438}.

\bibitem[{\citenamefont{Geng et~al.}(2022{\natexlab{a}})\citenamefont{Geng,
  Karch, Perez-Pardavila, Raju, Randall, Riojas, and
  Shashi}}]{Geng:2021entPhaseStruc}
\bibinfo{author}{\bibfnamefont{H.}~\bibnamefont{Geng}},
  \bibinfo{author}{\bibfnamefont{A.}~\bibnamefont{Karch}},
  \bibinfo{author}{\bibfnamefont{C.}~\bibnamefont{Perez-Pardavila}},
  \bibinfo{author}{\bibfnamefont{S.}~\bibnamefont{Raju}},
  \bibinfo{author}{\bibfnamefont{L.}~\bibnamefont{Randall}},
  \bibinfo{author}{\bibfnamefont{M.}~\bibnamefont{Riojas}}, \bibnamefont{and}
  \bibinfo{author}{\bibfnamefont{S.}~\bibnamefont{Shashi}},
  \bibinfo{journal}{JHEP} \textbf{\bibinfo{volume}{05}}, \bibinfo{pages}{153}
  (\bibinfo{year}{2022}{\natexlab{a}}), \eprint{2112.09132}.

\bibitem[{\citenamefont{Geng et~al.}(2021{\natexlab{a}})\citenamefont{Geng,
  Karch, Perez-Pardavila, Raju, Randall, Riojas, and
  Shashi}}]{Geng:2020infTransf}
\bibinfo{author}{\bibfnamefont{H.}~\bibnamefont{Geng}},
  \bibinfo{author}{\bibfnamefont{A.}~\bibnamefont{Karch}},
  \bibinfo{author}{\bibfnamefont{C.}~\bibnamefont{Perez-Pardavila}},
  \bibinfo{author}{\bibfnamefont{S.}~\bibnamefont{Raju}},
  \bibinfo{author}{\bibfnamefont{L.}~\bibnamefont{Randall}},
  \bibinfo{author}{\bibfnamefont{M.}~\bibnamefont{Riojas}}, \bibnamefont{and}
  \bibinfo{author}{\bibfnamefont{S.}~\bibnamefont{Shashi}},
  \bibinfo{journal}{SciPost Phys.} \textbf{\bibinfo{volume}{10}},
  \bibinfo{pages}{103} (\bibinfo{year}{2021}{\natexlab{a}}),
  \eprint{2012.04671}.

\bibitem[{\citenamefont{Geng et~al.}(2022{\natexlab{b}})\citenamefont{Geng,
  Karch, Perez-Pardavila, Raju, Randall, Riojas, and
  Shashi}}]{Geng:2021inconsistency}
\bibinfo{author}{\bibfnamefont{H.}~\bibnamefont{Geng}},
  \bibinfo{author}{\bibfnamefont{A.}~\bibnamefont{Karch}},
  \bibinfo{author}{\bibfnamefont{C.}~\bibnamefont{Perez-Pardavila}},
  \bibinfo{author}{\bibfnamefont{S.}~\bibnamefont{Raju}},
  \bibinfo{author}{\bibfnamefont{L.}~\bibnamefont{Randall}},
  \bibinfo{author}{\bibfnamefont{M.}~\bibnamefont{Riojas}}, \bibnamefont{and}
  \bibinfo{author}{\bibfnamefont{S.}~\bibnamefont{Shashi}},
  \bibinfo{journal}{JHEP} \textbf{\bibinfo{volume}{01}}, \bibinfo{pages}{182}
  (\bibinfo{year}{2022}{\natexlab{b}}), \eprint{2107.03390}.

\bibitem[{\citenamefont{Geng et~al.}(2021{\natexlab{b}})\citenamefont{Geng,
  Nomura, and Sun}}]{Geng:2021infPardeSitter}
\bibinfo{author}{\bibfnamefont{H.}~\bibnamefont{Geng}},
  \bibinfo{author}{\bibfnamefont{Y.}~\bibnamefont{Nomura}}, \bibnamefont{and}
  \bibinfo{author}{\bibfnamefont{H.-Y.} \bibnamefont{Sun}},
  \bibinfo{journal}{Phys. Rev. D} \textbf{\bibinfo{volume}{103}},
  \bibinfo{pages}{126004} (\bibinfo{year}{2021}{\natexlab{b}}),
  \eprint{2103.07477}.

\bibitem[{\citenamefont{Ahn et~al.}(2022)\citenamefont{Ahn, Bak, Jeong, Kim,
  and Sun}}]{Ahn:2021chg}
\bibinfo{author}{\bibfnamefont{B.}~\bibnamefont{Ahn}},
  \bibinfo{author}{\bibfnamefont{S.-E.} \bibnamefont{Bak}},
  \bibinfo{author}{\bibfnamefont{H.-S.} \bibnamefont{Jeong}},
  \bibinfo{author}{\bibfnamefont{K.-Y.} \bibnamefont{Kim}}, \bibnamefont{and}
  \bibinfo{author}{\bibfnamefont{Y.-W.} \bibnamefont{Sun}},
  \bibinfo{journal}{Phys. Rev. D} \textbf{\bibinfo{volume}{105}},
  \bibinfo{pages}{046012} (\bibinfo{year}{2022}), \eprint{2107.07444}.

\bibitem[{\citenamefont{Karananas et~al.}(2021)\citenamefont{Karananas,
  Kehagias, and Taskas}}]{Karananas:2021dilaton}
\bibinfo{author}{\bibfnamefont{G.~K.} \bibnamefont{Karananas}},
  \bibinfo{author}{\bibfnamefont{A.}~\bibnamefont{Kehagias}}, \bibnamefont{and}
  \bibinfo{author}{\bibfnamefont{J.}~\bibnamefont{Taskas}},
  \bibinfo{journal}{Journal of High Energy Physics}
  \textbf{\bibinfo{volume}{2021}} (\bibinfo{year}{2021}).

\bibitem[{\citenamefont{Hashimoto et~al.}(2020)\citenamefont{Hashimoto, Iizuka,
  and Matsuo}}]{Hashimoto_2020}
\bibinfo{author}{\bibfnamefont{K.}~\bibnamefont{Hashimoto}},
  \bibinfo{author}{\bibfnamefont{N.}~\bibnamefont{Iizuka}}, \bibnamefont{and}
  \bibinfo{author}{\bibfnamefont{Y.}~\bibnamefont{Matsuo}},
  \bibinfo{journal}{Journal of High Energy Physics}
  \textbf{\bibinfo{volume}{2020}} (\bibinfo{year}{2020}), \eprint{2004.05863}.

\bibitem[{\citenamefont{Ryu and
  Takayanagi}(2006{\natexlab{a}})}]{Ryu:2006holographic}
\bibinfo{author}{\bibfnamefont{S.}~\bibnamefont{Ryu}} \bibnamefont{and}
  \bibinfo{author}{\bibfnamefont{T.}~\bibnamefont{Takayanagi}},
  \bibinfo{journal}{Physical Review Letters} \textbf{\bibinfo{volume}{96}}
  (\bibinfo{year}{2006}{\natexlab{a}}).

\bibitem[{\citenamefont{Lewkowycz and
  Maldacena}(2013)}]{Lewkowycz:2013generalized}
\bibinfo{author}{\bibfnamefont{A.}~\bibnamefont{Lewkowycz}} \bibnamefont{and}
  \bibinfo{author}{\bibfnamefont{J.}~\bibnamefont{Maldacena}},
  \bibinfo{journal}{Journal of High Energy Physics}
  \textbf{\bibinfo{volume}{2013}} (\bibinfo{year}{2013}).

\bibitem[{\citenamefont{Engelhardt and
  Wall}(2015{\natexlab{a}})}]{Engelhardt:2015extremal}
\bibinfo{author}{\bibfnamefont{N.}~\bibnamefont{Engelhardt}} \bibnamefont{and}
  \bibinfo{author}{\bibfnamefont{A.~C.} \bibnamefont{Wall}},
  \bibinfo{journal}{Journal of High Energy Physics}
  \textbf{\bibinfo{volume}{2015}} (\bibinfo{year}{2015}{\natexlab{a}}).

\bibitem[{\citenamefont{Hubeny et~al.}(2007{\natexlab{a}})\citenamefont{Hubeny,
  Rangamani, and Takayanagi}}]{Hubeny:2007covariant}
\bibinfo{author}{\bibfnamefont{V.~E.} \bibnamefont{Hubeny}},
  \bibinfo{author}{\bibfnamefont{M.}~\bibnamefont{Rangamani}},
  \bibnamefont{and}
  \bibinfo{author}{\bibfnamefont{T.}~\bibnamefont{Takayanagi}},
  \bibinfo{journal}{Journal of High Energy Physics}
  \textbf{\bibinfo{volume}{2007}}, \bibinfo{pages}{062}
  (\bibinfo{year}{2007}{\natexlab{a}}).

\bibitem[{\citenamefont{Faulkner et~al.}(2013)\citenamefont{Faulkner,
  Lewkowycz, and Maldacena}}]{Faulkner:2013corrections}
\bibinfo{author}{\bibfnamefont{T.}~\bibnamefont{Faulkner}},
  \bibinfo{author}{\bibfnamefont{A.}~\bibnamefont{Lewkowycz}},
  \bibnamefont{and}
  \bibinfo{author}{\bibfnamefont{J.}~\bibnamefont{Maldacena}},
  \bibinfo{journal}{Journal of High Energy Physics}
  \textbf{\bibinfo{volume}{2013}} (\bibinfo{year}{2013}).

\bibitem[{\citenamefont{Penington et~al.}(2020)\citenamefont{Penington,
  Shenker, Stanford, and Yang}}]{penington:2020replica}
\bibinfo{author}{\bibfnamefont{G.}~\bibnamefont{Penington}},
  \bibinfo{author}{\bibfnamefont{S.~H.} \bibnamefont{Shenker}},
  \bibinfo{author}{\bibfnamefont{D.}~\bibnamefont{Stanford}}, \bibnamefont{and}
  \bibinfo{author}{\bibfnamefont{Z.}~\bibnamefont{Yang}}
  (\bibinfo{year}{2020}), \eprint{1911.11977}.

\bibitem[{\citenamefont{Almheiri
  et~al.}(2020{\natexlab{b}})\citenamefont{Almheiri, Hartman, Maldacena,
  Shaghoulian, and Tajdini}}]{Almheiri:2020wormholes}
\bibinfo{author}{\bibfnamefont{A.}~\bibnamefont{Almheiri}},
  \bibinfo{author}{\bibfnamefont{T.}~\bibnamefont{Hartman}},
  \bibinfo{author}{\bibfnamefont{J.}~\bibnamefont{Maldacena}},
  \bibinfo{author}{\bibfnamefont{E.}~\bibnamefont{Shaghoulian}},
  \bibnamefont{and} \bibinfo{author}{\bibfnamefont{A.}~\bibnamefont{Tajdini}},
  \bibinfo{journal}{Journal of High Energy Physics}
  \textbf{\bibinfo{volume}{2020}} (\bibinfo{year}{2020}{\natexlab{b}}).

\bibitem[{\citenamefont{Callan and
  Wilczek}(1994{\natexlab{a}})}]{Callan:1994geometric}
\bibinfo{author}{\bibfnamefont{C.}~\bibnamefont{Callan}} \bibnamefont{and}
  \bibinfo{author}{\bibfnamefont{F.}~\bibnamefont{Wilczek}},
  \bibinfo{journal}{Physics Letters B} \textbf{\bibinfo{volume}{333}},
  \bibinfo{pages}{55} (\bibinfo{year}{1994}{\natexlab{a}}).

\bibitem[{\citenamefont{Holzhey
  et~al.}(1994{\natexlab{a}})\citenamefont{Holzhey, Larsen, and
  Wilczek}}]{Holzhey:1994renormalized}
\bibinfo{author}{\bibfnamefont{C.}~\bibnamefont{Holzhey}},
  \bibinfo{author}{\bibfnamefont{F.}~\bibnamefont{Larsen}}, \bibnamefont{and}
  \bibinfo{author}{\bibfnamefont{F.}~\bibnamefont{Wilczek}},
  \bibinfo{journal}{Nuclear Physics B} \textbf{\bibinfo{volume}{424}},
  \bibinfo{pages}{443} (\bibinfo{year}{1994}{\natexlab{a}}).

\bibitem[{\citenamefont{Calabrese and
  Cardy}(2009)}]{Calabrese:2009entanglement}
\bibinfo{author}{\bibfnamefont{P.}~\bibnamefont{Calabrese}} \bibnamefont{and}
  \bibinfo{author}{\bibfnamefont{J.}~\bibnamefont{Cardy}},
  \bibinfo{journal}{Journal of Physics A: Mathematical and Theoretical}
  \textbf{\bibinfo{volume}{42}}, \bibinfo{pages}{504005}
  (\bibinfo{year}{2009}).

\bibitem[{\citenamefont{Chen et~al.}(2020{\natexlab{a}})\citenamefont{Chen,
  Fisher, Hernandez, Myers, and Ruan}}]{Chen:2020evaporation}
\bibinfo{author}{\bibfnamefont{H.~Z.} \bibnamefont{Chen}},
  \bibinfo{author}{\bibfnamefont{Z.}~\bibnamefont{Fisher}},
  \bibinfo{author}{\bibfnamefont{J.}~\bibnamefont{Hernandez}},
  \bibinfo{author}{\bibfnamefont{R.~C.} \bibnamefont{Myers}}, \bibnamefont{and}
  \bibinfo{author}{\bibfnamefont{S.-M.} \bibnamefont{Ruan}},
  \bibinfo{journal}{Journal of High Energy Physics}
  \textbf{\bibinfo{volume}{2020}} (\bibinfo{year}{2020}{\natexlab{a}}).

\bibitem[{\citenamefont{Chen}(2020)}]{Chen:2020pulling}
\bibinfo{author}{\bibfnamefont{Y.}~\bibnamefont{Chen}},
  \bibinfo{journal}{Journal of High Energy Physics}
  \textbf{\bibinfo{volume}{2020}} (\bibinfo{year}{2020}).

\bibitem[{\citenamefont{Akers et~al.}(2020)\citenamefont{Akers, Engelhardt,
  Penington, and Usatyuk}}]{akers:2020quantum}
\bibinfo{author}{\bibfnamefont{C.}~\bibnamefont{Akers}},
  \bibinfo{author}{\bibfnamefont{N.}~\bibnamefont{Engelhardt}},
  \bibinfo{author}{\bibfnamefont{G.}~\bibnamefont{Penington}},
  \bibnamefont{and} \bibinfo{author}{\bibfnamefont{M.}~\bibnamefont{Usatyuk}}
  (\bibinfo{year}{2020}), \eprint{1912.02799}.

\bibitem[{\citenamefont{Liu and Vardhan}(2021)}]{Liu:2021dynamical}
\bibinfo{author}{\bibfnamefont{H.}~\bibnamefont{Liu}} \bibnamefont{and}
  \bibinfo{author}{\bibfnamefont{S.}~\bibnamefont{Vardhan}},
  \bibinfo{journal}{Journal of High Energy Physics}
  \textbf{\bibinfo{volume}{2021}} (\bibinfo{year}{2021}).

\bibitem[{\citenamefont{Marolf and Maxfield}(2020)}]{Marolf:2020transcenting}
\bibinfo{author}{\bibfnamefont{D.}~\bibnamefont{Marolf}} \bibnamefont{and}
  \bibinfo{author}{\bibfnamefont{H.}~\bibnamefont{Maxfield}},
  \bibinfo{journal}{Journal of High Energy Physics}
  \textbf{\bibinfo{volume}{2020}} (\bibinfo{year}{2020}).

\bibitem[{\citenamefont{Balasubramanian
  et~al.}(2021)\citenamefont{Balasubramanian, Kar, Parrikar, S{\'{a}}rosi, and
  Ugajin}}]{Balasubramanian:2021secret}
\bibinfo{author}{\bibfnamefont{V.}~\bibnamefont{Balasubramanian}},
  \bibinfo{author}{\bibfnamefont{A.}~\bibnamefont{Kar}},
  \bibinfo{author}{\bibfnamefont{O.}~\bibnamefont{Parrikar}},
  \bibinfo{author}{\bibfnamefont{G.}~\bibnamefont{S{\'{a}}rosi}},
  \bibnamefont{and} \bibinfo{author}{\bibfnamefont{T.}~\bibnamefont{Ugajin}},
  \bibinfo{journal}{Journal of High Energy Physics}
  \textbf{\bibinfo{volume}{2021}} (\bibinfo{year}{2021}).

\bibitem[{\citenamefont{Bhattacharya}(2020)}]{Bhattacharya:2020purification}
\bibinfo{author}{\bibfnamefont{A.}~\bibnamefont{Bhattacharya}},
  \bibinfo{journal}{Physical Review D} \textbf{\bibinfo{volume}{102}}
  (\bibinfo{year}{2020}).

\bibitem[{\citenamefont{Verlinde}(2020)}]{verlinde:2020er}
\bibinfo{author}{\bibfnamefont{H.}~\bibnamefont{Verlinde}}
  (\bibinfo{year}{2020}), \eprint{2003.13117}.

\bibitem[{\citenamefont{Chen et~al.}(2020{\natexlab{b}})\citenamefont{Chen, Qi,
  and Zhang}}]{Chen:2020majorana}
\bibinfo{author}{\bibfnamefont{Y.}~\bibnamefont{Chen}},
  \bibinfo{author}{\bibfnamefont{X.-L.} \bibnamefont{Qi}}, \bibnamefont{and}
  \bibinfo{author}{\bibfnamefont{P.}~\bibnamefont{Zhang}},
  \bibinfo{journal}{Journal of High Energy Physics}
  \textbf{\bibinfo{volume}{2020}} (\bibinfo{year}{2020}{\natexlab{b}}).

\bibitem[{\citenamefont{Gautason et~al.}(2020)\citenamefont{Gautason,
  Schneiderbauer, Sybesma, and Thorlacius}}]{Gautason:2020page}
\bibinfo{author}{\bibfnamefont{F.~F.} \bibnamefont{Gautason}},
  \bibinfo{author}{\bibfnamefont{L.}~\bibnamefont{Schneiderbauer}},
  \bibinfo{author}{\bibfnamefont{W.}~\bibnamefont{Sybesma}}, \bibnamefont{and}
  \bibinfo{author}{\bibfnamefont{L.}~\bibnamefont{Thorlacius}},
  \bibinfo{journal}{Journal of High Energy Physics}
  \textbf{\bibinfo{volume}{2020}} (\bibinfo{year}{2020}).

\bibitem[{\citenamefont{Anegawa and Iizuka}(2020)}]{Anegawa:2020notes}
\bibinfo{author}{\bibfnamefont{T.}~\bibnamefont{Anegawa}} \bibnamefont{and}
  \bibinfo{author}{\bibfnamefont{N.}~\bibnamefont{Iizuka}},
  \bibinfo{journal}{Journal of High Energy Physics}
  \textbf{\bibinfo{volume}{2020}} (\bibinfo{year}{2020}).

\bibitem[{\citenamefont{Almheiri
  et~al.}(2020{\natexlab{c}})\citenamefont{Almheiri, Mahajan, and
  Santos}}]{Almheiri:2020higher}
\bibinfo{author}{\bibfnamefont{A.}~\bibnamefont{Almheiri}},
  \bibinfo{author}{\bibfnamefont{R.}~\bibnamefont{Mahajan}}, \bibnamefont{and}
  \bibinfo{author}{\bibfnamefont{J.}~\bibnamefont{Santos}},
  \bibinfo{journal}{{SciPost} Physics} \textbf{\bibinfo{volume}{9}}
  (\bibinfo{year}{2020}{\natexlab{c}}).

\bibitem[{\citenamefont{Schwarzschild}(1999)}]{schwarzschild:1999gravitational}
\bibinfo{author}{\bibfnamefont{K.}~\bibnamefont{Schwarzschild}}
  (\bibinfo{year}{1999}), \eprint{physics/9905030}.

\bibitem[{\citenamefont{Bardeen}(1968)}]{bardeen1968proceedings}
\bibinfo{author}{\bibfnamefont{J.~M.} \bibnamefont{Bardeen}},
  \emph{\bibinfo{title}{Proceedings of the international conference gr5}}
  (\bibinfo{year}{1968}).

\bibitem[{\citenamefont{Ayon-Beato and Garcia}(2000)}]{Ayon-Beato:2000mjt}
\bibinfo{author}{\bibfnamefont{E.}~\bibnamefont{Ayon-Beato}} \bibnamefont{and}
  \bibinfo{author}{\bibfnamefont{A.}~\bibnamefont{Garcia}},
  \bibinfo{journal}{Phys. Lett. B} \textbf{\bibinfo{volume}{493}},
  \bibinfo{pages}{149} (\bibinfo{year}{2000}), \eprint{gr-qc/0009077}.

\bibitem[{\citenamefont{Ay\'on-Beato and
  Garc\'{\i}a}(1998)}]{ayonbeato:1998regular}
\bibinfo{author}{\bibfnamefont{E.}~\bibnamefont{Ay\'on-Beato}}
  \bibnamefont{and}
  \bibinfo{author}{\bibfnamefont{A.}~\bibnamefont{Garc\'{\i}a}},
  \bibinfo{journal}{Phys. Rev. Lett.} \textbf{\bibinfo{volume}{80}},
  \bibinfo{pages}{5056} (\bibinfo{year}{1998}).

\bibitem[{\citenamefont{Hayward}(2006)}]{Hayward:2006nonsingular}
\bibinfo{author}{\bibfnamefont{S.~A.} \bibnamefont{Hayward}},
  \bibinfo{journal}{Phys. Rev. Lett.} \textbf{\bibinfo{volume}{96}},
  \bibinfo{pages}{031103} (\bibinfo{year}{2006}).

\bibitem[{\citenamefont{Ryu and Takayanagi}(2006{\natexlab{b}})}]{Ryu:2006bv}
\bibinfo{author}{\bibfnamefont{S.}~\bibnamefont{Ryu}} \bibnamefont{and}
  \bibinfo{author}{\bibfnamefont{T.}~\bibnamefont{Takayanagi}},
  \bibinfo{journal}{Phys. Rev. Lett.} \textbf{\bibinfo{volume}{96}},
  \bibinfo{pages}{181602} (\bibinfo{year}{2006}{\natexlab{b}}),
  \eprint{hep-th/0603001}.

\bibitem[{\citenamefont{Hubeny et~al.}(2007{\natexlab{b}})\citenamefont{Hubeny,
  Rangamani, and Takayanagi}}]{Hubeny:2007xt}
\bibinfo{author}{\bibfnamefont{V.~E.} \bibnamefont{Hubeny}},
  \bibinfo{author}{\bibfnamefont{M.}~\bibnamefont{Rangamani}},
  \bibnamefont{and}
  \bibinfo{author}{\bibfnamefont{T.}~\bibnamefont{Takayanagi}},
  \bibinfo{journal}{JHEP} \textbf{\bibinfo{volume}{07}}, \bibinfo{pages}{062}
  (\bibinfo{year}{2007}{\natexlab{b}}), \eprint{0705.0016}.

\bibitem[{\citenamefont{Engelhardt and
  Wall}(2015{\natexlab{b}})}]{Engelhardt:2014gca}
\bibinfo{author}{\bibfnamefont{N.}~\bibnamefont{Engelhardt}} \bibnamefont{and}
  \bibinfo{author}{\bibfnamefont{A.~C.} \bibnamefont{Wall}},
  \bibinfo{journal}{JHEP} \textbf{\bibinfo{volume}{01}}, \bibinfo{pages}{073}
  (\bibinfo{year}{2015}{\natexlab{b}}), \eprint{1408.3203}.

\bibitem[{\citenamefont{Callan and
  Wilczek}(1994{\natexlab{b}})}]{Callan:1994py}
\bibinfo{author}{\bibfnamefont{C.~G.} \bibnamefont{Callan}, \bibfnamefont{Jr.}}
  \bibnamefont{and} \bibinfo{author}{\bibfnamefont{F.}~\bibnamefont{Wilczek}},
  \bibinfo{journal}{Phys. Lett. B} \textbf{\bibinfo{volume}{333}},
  \bibinfo{pages}{55} (\bibinfo{year}{1994}{\natexlab{b}}),
  \eprint{hep-th/9401072}.

\bibitem[{\citenamefont{Holzhey
  et~al.}(1994{\natexlab{b}})\citenamefont{Holzhey, Larsen, and
  Wilczek}}]{Holzhey:1994we}
\bibinfo{author}{\bibfnamefont{C.}~\bibnamefont{Holzhey}},
  \bibinfo{author}{\bibfnamefont{F.}~\bibnamefont{Larsen}}, \bibnamefont{and}
  \bibinfo{author}{\bibfnamefont{F.}~\bibnamefont{Wilczek}},
  \bibinfo{journal}{Nucl. Phys. B} \textbf{\bibinfo{volume}{424}},
  \bibinfo{pages}{443} (\bibinfo{year}{1994}{\natexlab{b}}),
  \eprint{hep-th/9403108}.

\bibitem[{\citenamefont{Penington et~al.}(2022)\citenamefont{Penington,
  Shenker, Stanford, and Yang}}]{Penington:2019kki}
\bibinfo{author}{\bibfnamefont{G.}~\bibnamefont{Penington}},
  \bibinfo{author}{\bibfnamefont{S.~H.} \bibnamefont{Shenker}},
  \bibinfo{author}{\bibfnamefont{D.}~\bibnamefont{Stanford}}, \bibnamefont{and}
  \bibinfo{author}{\bibfnamefont{Z.}~\bibnamefont{Yang}},
  \bibinfo{journal}{JHEP} \textbf{\bibinfo{volume}{03}}, \bibinfo{pages}{205}
  (\bibinfo{year}{2022}), \eprint{1911.11977}.

\bibitem[{\citenamefont{Hartman et~al.}(2020)\citenamefont{Hartman,
  Shaghoulian, and Strominger}}]{Hartman:2020swn}
\bibinfo{author}{\bibfnamefont{T.}~\bibnamefont{Hartman}},
  \bibinfo{author}{\bibfnamefont{E.}~\bibnamefont{Shaghoulian}},
  \bibnamefont{and}
  \bibinfo{author}{\bibfnamefont{A.}~\bibnamefont{Strominger}},
  \bibinfo{journal}{JHEP} \textbf{\bibinfo{volume}{07}}, \bibinfo{pages}{022}
  (\bibinfo{year}{2020}), \eprint{2004.13857}.

\bibitem[{\citenamefont{Goto et~al.}(2021)\citenamefont{Goto, Hartman, and
  Tajdini}}]{Goto:2020wnk}
\bibinfo{author}{\bibfnamefont{K.}~\bibnamefont{Goto}},
  \bibinfo{author}{\bibfnamefont{T.}~\bibnamefont{Hartman}}, \bibnamefont{and}
  \bibinfo{author}{\bibfnamefont{A.}~\bibnamefont{Tajdini}},
  \bibinfo{journal}{JHEP} \textbf{\bibinfo{volume}{04}}, \bibinfo{pages}{289}
  (\bibinfo{year}{2021}), \eprint{2011.09043}.

\bibitem[{\citenamefont{Bambi et~al.}(2014)\citenamefont{Bambi, Malafarina, and
  Tsukamoto}}]{Bambi:2014accretion}
\bibinfo{author}{\bibfnamefont{C.}~\bibnamefont{Bambi}},
  \bibinfo{author}{\bibfnamefont{D.}~\bibnamefont{Malafarina}},
  \bibnamefont{and}
  \bibinfo{author}{\bibfnamefont{N.}~\bibnamefont{Tsukamoto}},
  \bibinfo{journal}{Physical Review D} \textbf{\bibinfo{volume}{89}}
  (\bibinfo{year}{2014}).

\bibitem[{\citenamefont{Bambi and Modesto}(2013)}]{Bambi:2013ufa}
\bibinfo{author}{\bibfnamefont{C.}~\bibnamefont{Bambi}} \bibnamefont{and}
  \bibinfo{author}{\bibfnamefont{L.}~\bibnamefont{Modesto}},
  \bibinfo{journal}{Phys. Lett. B} \textbf{\bibinfo{volume}{721}},
  \bibinfo{pages}{329} (\bibinfo{year}{2013}), \eprint{1302.6075}.

\bibitem[{\citenamefont{Newman and Janis}(2004)}]{NewmanJanis:1965algorithm}
\bibinfo{author}{\bibfnamefont{E.~T.} \bibnamefont{Newman}} \bibnamefont{and}
  \bibinfo{author}{\bibfnamefont{A.~I.} \bibnamefont{Janis}},
  \bibinfo{journal}{Journal of Mathematical Physics}
  \textbf{\bibinfo{volume}{6}}, \bibinfo{pages}{915} (\bibinfo{year}{2004}),
  ISSN \bibinfo{issn}{0022-2488}.

\bibitem[{\citenamefont{Newman et~al.}(2004)\citenamefont{Newman, Couch,
  Chinnapared, Exton, Prakash, and Torrence}}]{Newman:1965algorithmcharged}
\bibinfo{author}{\bibfnamefont{E.~T.} \bibnamefont{Newman}},
  \bibinfo{author}{\bibfnamefont{E.}~\bibnamefont{Couch}},
  \bibinfo{author}{\bibfnamefont{K.}~\bibnamefont{Chinnapared}},
  \bibinfo{author}{\bibfnamefont{A.}~\bibnamefont{Exton}},
  \bibinfo{author}{\bibfnamefont{A.}~\bibnamefont{Prakash}}, \bibnamefont{and}
  \bibinfo{author}{\bibfnamefont{R.}~\bibnamefont{Torrence}},
  \bibinfo{journal}{Journal of Mathematical Physics}
  \textbf{\bibinfo{volume}{6}}, \bibinfo{pages}{918} (\bibinfo{year}{2004}),
  ISSN \bibinfo{issn}{0022-2488}.

\bibitem[{\citenamefont{Wang et~al.}(2021)\citenamefont{Wang, Li, and
  Wang}}]{Wang_2021}
\bibinfo{author}{\bibfnamefont{X.}~\bibnamefont{Wang}},
  \bibinfo{author}{\bibfnamefont{R.}~\bibnamefont{Li}}, \bibnamefont{and}
  \bibinfo{author}{\bibfnamefont{J.}~\bibnamefont{Wang}},
  \bibinfo{journal}{Journal of High Energy Physics}
  \textbf{\bibinfo{volume}{2021}} (\bibinfo{year}{2021}), \eprint{2101.06867}.

\bibitem[{\citenamefont{Mansoori et~al.}(2022)\citenamefont{Mansoori, Luongo,
  Mancini, Mirjalali, Rafiee, and Tavanfar}}]{Mansoori_etall:2022}
\bibinfo{author}{\bibfnamefont{S.~A.~H.} \bibnamefont{Mansoori}},
  \bibinfo{author}{\bibfnamefont{O.}~\bibnamefont{Luongo}},
  \bibinfo{author}{\bibfnamefont{S.}~\bibnamefont{Mancini}},
  \bibinfo{author}{\bibfnamefont{M.}~\bibnamefont{Mirjalali}},
  \bibinfo{author}{\bibfnamefont{M.}~\bibnamefont{Rafiee}}, \bibnamefont{and}
  \bibinfo{author}{\bibfnamefont{A.}~\bibnamefont{Tavanfar}},
  \bibinfo{journal}{Physical Review D} \textbf{\bibinfo{volume}{106}}
  (\bibinfo{year}{2022}).

\bibitem[{\citenamefont{Boshkayev et~al.}(2023)\citenamefont{Boshkayev,
  Idrissov, Luongo, and Muccino}}]{orl1}
\bibinfo{author}{\bibfnamefont{K.}~\bibnamefont{Boshkayev}},
  \bibinfo{author}{\bibfnamefont{A.}~\bibnamefont{Idrissov}},
  \bibinfo{author}{\bibfnamefont{O.}~\bibnamefont{Luongo}}, \bibnamefont{and}
  \bibinfo{author}{\bibfnamefont{M.}~\bibnamefont{Muccino}}
  (\bibinfo{year}{2023}), \eprint{2303.03248}.

\bibitem[{\citenamefont{Malafarina}(2022)}]{Malafarina:2022wmx}
\bibinfo{author}{\bibfnamefont{D.}~\bibnamefont{Malafarina}}
  (\bibinfo{year}{2022}), \eprint{2209.11406}.

\bibitem[{\citenamefont{Zhou and Modesto}(2023)}]{Zhou:2023analyticRBH}
\bibinfo{author}{\bibfnamefont{T.}~\bibnamefont{Zhou}} \bibnamefont{and}
  \bibinfo{author}{\bibfnamefont{L.}~\bibnamefont{Modesto}}
  (\bibinfo{year}{2023}), \eprint{2303.11322}.

\bibitem[{\citenamefont{Giacchini et~al.}(2021)\citenamefont{Giacchini,
  de~Paula~Netto, and Modesto}}]{Giacchini:2021highrOrderDerivative}
\bibinfo{author}{\bibfnamefont{B.~L.} \bibnamefont{Giacchini}},
  \bibinfo{author}{\bibfnamefont{T.}~\bibnamefont{de~Paula~Netto}},
  \bibnamefont{and} \bibinfo{author}{\bibfnamefont{L.}~\bibnamefont{Modesto}},
  \bibinfo{journal}{Phys. Rev. D} \textbf{\bibinfo{volume}{104}},
  \bibinfo{pages}{084072} (\bibinfo{year}{2021}).

\bibitem[{\citenamefont{Kim and Nam}(2021)}]{Kim:2021entanglement}
\bibinfo{author}{\bibfnamefont{W.}~\bibnamefont{Kim}} \bibnamefont{and}
  \bibinfo{author}{\bibfnamefont{M.}~\bibnamefont{Nam}}, \bibinfo{journal}{The
  European Physical Journal C} \textbf{\bibinfo{volume}{81}}
  (\bibinfo{year}{2021}).

\bibitem[{\citenamefont{Saha et~al.}(2022)\citenamefont{Saha, Gangopadhyay, and
  Saha}}]{Saha:2021mutualInfIsland}
\bibinfo{author}{\bibfnamefont{A.}~\bibnamefont{Saha}},
  \bibinfo{author}{\bibfnamefont{S.}~\bibnamefont{Gangopadhyay}},
  \bibnamefont{and} \bibinfo{author}{\bibfnamefont{J.~P.} \bibnamefont{Saha}},
  \bibinfo{journal}{Eur. Phys. J. C} \textbf{\bibinfo{volume}{82}},
  \bibinfo{pages}{476} (\bibinfo{year}{2022}), \eprint{2109.02996}.

\bibitem[{\citenamefont{Roy~Chowdhury et~al.}(2022)\citenamefont{Roy~Chowdhury,
  Saha, and Gangopadhyay}}]{RoyChowdhury:2022mutualInfPage}
\bibinfo{author}{\bibfnamefont{A.}~\bibnamefont{Roy~Chowdhury}},
  \bibinfo{author}{\bibfnamefont{A.}~\bibnamefont{Saha}}, \bibnamefont{and}
  \bibinfo{author}{\bibfnamefont{S.}~\bibnamefont{Gangopadhyay}},
  \bibinfo{journal}{Phys. Rev. D} \textbf{\bibinfo{volume}{106}},
  \bibinfo{pages}{086019} (\bibinfo{year}{2022}), \eprint{2207.13029}.

\bibitem[{\citenamefont{Roy~Chowdhury et~al.}(2023)\citenamefont{Roy~Chowdhury,
  Saha, and Gangopadhyay}}]{RoyChowdhury:2023mutualSchdeSit}
\bibinfo{author}{\bibfnamefont{A.}~\bibnamefont{Roy~Chowdhury}},
  \bibinfo{author}{\bibfnamefont{A.}~\bibnamefont{Saha}}, \bibnamefont{and}
  \bibinfo{author}{\bibfnamefont{S.}~\bibnamefont{Gangopadhyay}}
  (\bibinfo{year}{2023}), \eprint{2303.14062}.

\bibitem[{\citenamefont{Il'ich~Kruglov}(2021)}]{IlichKruglov:2021pdw}
\bibinfo{author}{\bibfnamefont{S.}~\bibnamefont{Il'ich~Kruglov}},
  \bibinfo{journal}{Grav. Cosmol.} \textbf{\bibinfo{volume}{27}},
  \bibinfo{pages}{78} (\bibinfo{year}{2021}), \eprint{2103.14087}.

\bibitem[{\citenamefont{Wald}(2001)}]{Wald:1999vt}
\bibinfo{author}{\bibfnamefont{R.~M.} \bibnamefont{Wald}},
  \bibinfo{journal}{Living Rev. Rel.} \textbf{\bibinfo{volume}{4}},
  \bibinfo{pages}{6} (\bibinfo{year}{2001}), \eprint{gr-qc/9912119}.

\bibitem[{\citenamefont{Luongo and Quevedo}(2023)}]{Luongo:2023aib}
\bibinfo{author}{\bibfnamefont{O.}~\bibnamefont{Luongo}} \bibnamefont{and}
  \bibinfo{author}{\bibfnamefont{H.}~\bibnamefont{Quevedo}}
  (\bibinfo{year}{2023}), \eprint{2305.11185}.

\end{thebibliography}
\end{document}